\documentclass[12pt]{article}

\usepackage{latexsym}
\usepackage{amssymb}
\usepackage{amsbsy}
\usepackage{amsmath}
\usepackage{epsfig}
\usepackage{graphics,color}
\usepackage{a4}

\newcommand{\rmR}{{\rm R}}
\newcommand{\rmI}{{\rm I}}
\newcommand{\rmMax}{{\rm max}}

\newcommand{\be}{\begin{equation}}
\newcommand{\ee}{\end{equation}}
\newcommand{\bea}{\begin{eqnarray}}
\newcommand{\eea}{\end{eqnarray}}
\newcommand{\bean}{\begin{eqnarray*}}
\newcommand{\eean}{\end{eqnarray*}}
\newcommand{\nn}{\nonumber}
\newcommand{\hm}{\hspace*{-0.6cm}}
\newcommand{\bra}{\langle}
\newcommand{\ket}{\rangle}
\newcommand{\eps}{\epsilon}

\newcommand{\Tr}{\mbox{Tr}\,}
\newcommand{\xv}{{\mathbf x}}
\newcommand{\cP}{{\cal P}}
\newcommand{\vecx}{{\mathbf x}}

\begin{document}
\title{ 
\vskip -100pt
{
\begin{normalsize}
\mbox{} \hfill \\
\mbox{} \hfill arXiv:0912.3360 [hep-lat]
\vskip  70pt
\end{normalsize}
}
\bf\large
Adaptive stepsize and instabilities \\
in complex Langevin dynamics
}

\author{
\addtocounter{footnote}{2}
 Gert Aarts$^a$\thanks{email: g.aarts@swan.ac.uk} \,\,\,\,
 Frank A.\ James$^a$\thanks{email: pyfj@swan.ac.uk} \,\,\,\,
\addtocounter{footnote}{1}
 Erhard Seiler$^b$\thanks{email: ehs@mppmu.mpg.de} 
\\ and
 Ion-Olimpiu Stamatescu$^c$\thanks{email: 
I.O.Stamatescu@thphys.uni-heidelberg.de} \\ 
\mbox{} \\
 {$^a$\em\normalsize Department of Physics, Swansea University} \\
 {\em\normalsize Swansea, United Kingdom} \\ 
 \mbox{} \\
 $^b${\em\normalsize Max-Planck-Institut f\"ur Physik
(Werner-Heisenberg-Institut)} \\
 {\em\normalsize M{\"u}nchen, Germany} \\
 \mbox{} \\
 $^c${\em\normalsize Institut f\"ur Theoretische Physik, Universit\"at 
Heidelberg and FEST} \\
 {\em\normalsize Heidelberg, Germany} \\
}

\date{December 3, 2009}
\maketitle

\begin{abstract}
 Stochastic quantization offers the opportunity to simulate field theories 
with a complex action. In some theories unstable trajectories are 
prevalent when a constant stepsize is employed. We construct algorithms 
for generating an adaptive stepsize in complex Langevin simulations and 
find that unstable trajectories are completely eliminated. To illustrate 
the generality of the approach, we apply it to the three-dimensional XY 
model at nonzero chemical potential and the heavy dense limit of QCD.
 \end{abstract}
\newpage

\section{Introduction}
\label{sec:Introduction}
\setcounter{equation}{0}

 Nonperturbative simulations of field theories with a complex action are 
difficult, since importance sampling based techniques break down.
 Stochastic quantisation and complex Langevin dynamics 
\cite{Parisi:1980ys,Parisi:1984cs,Klauder:1983,Damgaard:1987rr} can 
potentially evade this problem, as it is not based on a probability 
interpretation of the weight in the path integral. Among others, this is 
relevant for theories with a sign problem due to a nonzero chemical 
potential \cite{Karsch:1985cb,Aarts:2008rr,Aarts:2008wh,Aarts:2009hn} and for the 
dynamics of quantum fields out of equilibrium 
\cite{Berges:2005yt,Berges:2006xc,Berges:2007nr}. Other recent activity 
can be found in Refs.\ 
\cite{Bernard:2001wh,Pehlevan:2007eq}.

As is well known since the 80s, there are a number of problems associated 
with complex Langevin dynamics, see e.g.\ Refs.\ 
\cite{Klauder:1985b,Ambjorn:1985iw,Ambjorn:1986fz}. These can roughly be 
divided under two headings: instabilities and convergence. The first 
problem concerns the appearance of instabilities when solving the 
discretized Langevin equations numerically. Sometimes, but not always, 
this can be controlled by choosing a small enough stepsize. 
 The second problem pertains to convergence. In some cases complex 
Langevin simulations appear to converge but not to the correct answer 
(see e.g.\ Ref.\ \cite{Ambjorn:1986fz}). In order to disentangle these 
issues, we 
tackle in this paper the first one and present adaptive stepsize 
algorithms that lead to a stable evolution and are not constrained to very 
small stepsizes only. A discussion of the second problem is deferred to 
future publications.

The paper is organized as follows. In Sec.\ \ref{sec:adap} we introduce 
the problem, indicate why an adaptive stepsize might be necessary and 
outline the basic idea behind the algorithms, extending the ideas of 
Refs.\ \cite{Ambjorn:1985iw,Ambjorn:1986fz}. To avoid notational 
cluttering we use a real scalar field, but we emphasize that the method is 
more generally applicable. We then present two algorithms implementing the 
basic idea, and apply them to the three-dimensional XY model at finite 
chemical potential in Sec.~\ref{sec:xy} and the heavy dense limit of QCD 
in Sec.~\ref{sec:hdm}. The latter has previously been studied with complex 
Langevin dynamics in Ref.\ \cite{Aarts:2008rr}. We show a few selected 
results to indicate the applicability of the approach. As mentioned above, 
convergence will be discussed elsewhere.

\section{Adaptive stepsize}
\label{sec:adap}
\setcounter{equation}{0}

Consider a real scalar field $\phi$ with the Langevin equation of motion
 \be
 \frac{\partial \phi}{\partial \vartheta} = -\frac{\delta S}{\delta \phi} 
 + \eta.
 \ee
 Here $\vartheta$ is the supplementary Langevin time, $-\delta 
S/\delta\phi$ 
is the drift term derived from the action $S$, and $\eta$ is Gaussian 
noise. The fundamental assertion of stochastic quantisation is that in the 
infinite (Langevin) time limit, noise averages of observables become equal 
to quantum expectation values, defined via the standard functional 
integral,
 \be
\lim_{\vartheta\to\infty} \left\bra O[\phi(\vartheta)] \right\ket_\eta = 
\frac{\int D\phi \, e^{-S} O[\phi]}{\int D\phi \, e^{-S} },
 \ee
 where the brackets on the left denote a noise average.

If the action is complex the drift term becomes complex and so the field 
acquires an imaginary part (even if initially real). One must therefore 
complexify all fields, $\phi \to \phi^\rmR + i\phi^\rmI$. The 
Langevin equation then becomes
 \bea
\frac{\partial \phi^\rmR}{\partial \vartheta} = K^\rmR +\eta, 
\;\;\;\;\;\;\;\;\;\;\;\;\;
&& 
 K^\rmR = -\mbox{Re}\, \frac{\delta S}{\delta \phi}
	\Big|_{\phi\rightarrow\phi^\rmR+i\phi^\rmI}, 
\\
\frac{\partial \phi^\rmI}{\partial \vartheta} = K^\rmI,
\;\;\;\;\;\;\;\;\;\;\;\;\;\;\;\;\;\;\;\;
&& 
 K^\rmI = -\mbox{Im}\, \frac{\delta S}{\delta \phi}
	\Big|_{\phi\rightarrow\phi^\rmR+i\phi^\rmI}.
\eea
 Here we restricted ourselves to real noise.

\begin{figure}[t]
\begin{center}
\epsfig{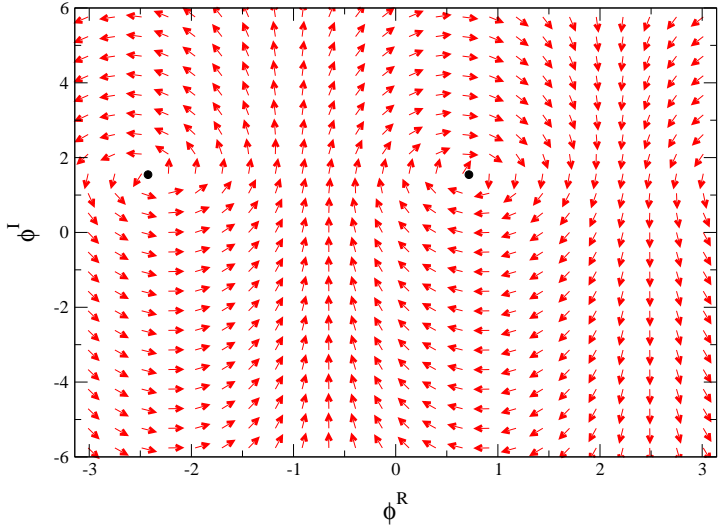}
\end{center}
 \caption{Example of a classical flow diagram in the XY model at 
nonzero chemical potential ($\mu=2$). The arrows denote the normalized 
drift terms ($K^\rmR,K^\rmI)$ at ($\phi^\rmR,\phi^\rmI$). The dots are 
classical fixed points.}
 \label{fig:flow} 
\end{figure}

 The complexification changes the dynamics substantially. Suppose that 
before complexification $\phi$ is a variable with a compact domain, 
e.g.\ $-\pi<\phi\leq \pi$. After complexification, the domain is 
noncompact since $-\infty<\phi^\rmI<\infty$. Moreover, there will be 
unstable directions along which $\phi^\rmI\to \pm\infty$. This is best 
seen in classical flow diagrams, in which the drift terms are plotted 
as a function of the degrees of freedom $\phi^\rmR, \phi^\rmI$.
 In Fig.\ \ref{fig:flow} we show an example of a classical flow diagram in 
the XY model at finite chemical potential (to be discussed below). More 
examples of classical flow diagrams with unstable directions can be found 
in Refs.\ 
\cite{Aarts:2008rr,Berges:2007nr,Pehlevan:2007eq,Ambjorn:1986fz}.
 The arrows denote the drift terms ($K^\rmR,K^\rmI)$ at 
($\phi^\rmR,\phi^\rmI$). The length of the arrows is normalized for 
clarity. In this case there are unstable directions at $\phi^\rmR \sim 
-0.7$ and $\phi^\rmR \sim 2.4$. The black dots denote classical fixed 
points where the drift terms vanish. Generally speaking, the forces are 
larger when one is further away from the fixed points. In absence of the 
noise, one finds generically that configurations reach infinity in a 
finite time, since the forces grow exponentially for large imaginary field 
values.

When a Langevin trajectory makes a large excursion into imaginary 
directions, for instance by coming close to an unstable direction, 
sufficient care in the numerical integration of the Langevin equations is 
mandatory. In some cases it suffices to employ a small stepsize $\eps$, 
after discretizing Langevin time as $\vartheta=n\eps$. However, this does 
not 
solve instabilities in all situations. Moreover, a small stepsize will 
result in a slow evolution, requiring many updates to explore 
configuration space.

To cure both problems, we introduce an adaptive stepsize, $\eps_n$, in the 
discretized Langevin equations,
 \begin{align}
 \phi_x^\rmR(n+1) &= \phi_x^\rmR(n) + \epsilon_n K^\rmR_x(n) + 
 \sqrt{\epsilon_n}\eta_x(n), \\
 \phi_x^\rmI(n+1) &= \phi_x^\rmI(n) + \epsilon_n K^\rmI_x(n),
 \label{eq:update}
 \end{align}
 where the noise satisfies
 \be
 \bra \eta_x(n)\ket = 0,
 \;\;\;\;\;\;\;\;\;\;\;\;
 \bra \eta_x(n)\eta_{x'}(n')\ket = 2\delta_{xx'}\delta_{nn'}.
 \ee
 The magnitude of the stepsize is determined by controlling the distance a 
single update makes in the configuration space. Here we present two 
specific algorithms to do this.

In the first formulation, we monitor, at each discrete Langevin time $n$, 
the quantity
 \be
 K^\rmMax_n = \max_{x}|K_x(n)| 
 = \max_{x} \sqrt{{K^\rmR_x}^2(n) + {K^\rmI_x}^2(n)}.
\ee
 We then place an upper bound on the product $\epsilon K^\rmMax$
and define the stepsize $\epsilon_n$ as
 \be
 \epsilon_n = \bar{\epsilon}\frac{\langle K^\rmMax \rangle}{K^\rmMax_n}.
 \ee
 Here $\bar{\epsilon}$ is the desired average stepsize (which can be 
controlled) and the expectation value of the maximum drift term $\langle 
K^\rmMax \rangle$ is either precomputed, or computed during the 
thermalisation phase (with an initial guess). In this way, the stepsize is 
completely local in Langevin time and becomes smaller when the drift term 
is large (e.g.\ close to an instability) and larger when it is safe to do 
so.

In the second formulation, we keep $\eps K^{\rmMax}$ within a factor $p$ 
relative to a reference value ${\cal K}$, i.e.,
 \be
 \frac{1}{p}\, {\cal K} \leq \eps {K}^{\rmMax} 
\leq p\,{\cal K}.
 \ee
 If this range is exceeded the stepsize is increased/reduced by the factor 
$p$. This is iterated several times, if necessary. Both $p$ and 
${\cal K}$ have to be chosen beforehand, but this 
does not 
require fine tuning as long as clearly inadequate regions are avoided.

In the next sections, we apply these formulations to the XY model at 
nonzero chemical potential and QCD in the heavy dense limit 
respectively.

\section{XY model}
\label{sec:xy}
\setcounter{equation}{0}

We demonstrate the first implementation using the three-dimensional XY 
model at finite chemical potential \cite{Chandrasekharan:2008gp}. The 
action is
 \be
 S = -\beta\sum_{x}\sum_{\nu=0}^2 \cos\left(\phi_{x} - \phi_{x+\hat\nu} - 
i\mu\delta_{\nu,0}\right).
 \ee
 The theory is defined on a lattice of size $\Omega=N_\tau N_s^2$, with 
periodic boundary conditions in all three directions. The chemical 
potential is introduced as an imaginary constant vector field in the 
temporal direction \cite{Hasenfratz:1983ba} and couples to the conserved 
Noether charge associated with the global symmetry 
$\phi_x\to\phi_x+\alpha$. As always, the action is complex when $\mu\neq 
0$ and satisfies $S^*(\mu)=S(-\mu^*)$.\footnote{At nonzero chemical 
potential, it is preferable to interpret this system as a 
three-dimensional euclidean quantum field theory at finite temperature 
(with coupling $\beta$ and inverse temperature $N_\tau$), rather than as a 
three-dimensional classical spin system with inverse temperature $\beta$.
Models in this class can also be studied using world-line
formulations \cite{Endres:2006xu,Chandrasekharan:2008gp}.}

The drift terms, after complexification, read
\bea
 K_x^\rmR =  -\beta\sum_\nu \Big[
 \sin\left(\phi^\rmR_x-\phi^\rmR_{x+\hat \nu}\right)
 \cosh\left(\phi^\rmI_x-\phi^\rmI_{x+\hat \nu}-\mu\delta_{\nu,0}\right)
&&\nn \\
 +\sin\left(\phi^\rmR_x-\phi^\rmR_{x-\hat \nu}\right)
 \cosh\left(\phi^\rmI_x-\phi^\rmI_{x-\hat \nu}+\mu\delta_{\nu,0}\right)
\Big],
&&
\\
 K_x^\rmI = -\beta\sum_\nu \Big[
 \cos\left(\phi^\rmR_x-\phi^\rmR_{x+\hat \nu}\right)
 \sinh\left(\phi^\rmI_x-\phi^\rmI_{x+\hat \nu}-\mu\delta_{\nu,0}\right)
&&\nn \\
+\cos\left(\phi^\rmR_x-\phi^\rmR_{x-\hat \nu}\right)
 \sinh\left(\phi^\rmI_x-\phi^\rmI_{x-\hat \nu}+\mu\delta_{\nu,0}\right)
\Big].
&&
\eea
 As anticipated, they are unbounded due to the $\phi^\rmI$ variables.

To construct the flow diagram in Fig.\ \ref{fig:flow}, we have chosen 
the field variables at the six sites neighbouring $\phi_x$ as random 
variables between $\pm \pi$. Note that the drift terms change sign when 
$\phi_x\to \phi_x+\pi$, for given $x$, explaining the symmetry in Fig.\ 
\ref{fig:flow}. The normalized drift terms and the classical fixed points 
($K_x^\rmR=K_x^\rmI=0$) are independent of $\beta$.

\begin{figure}[p!]
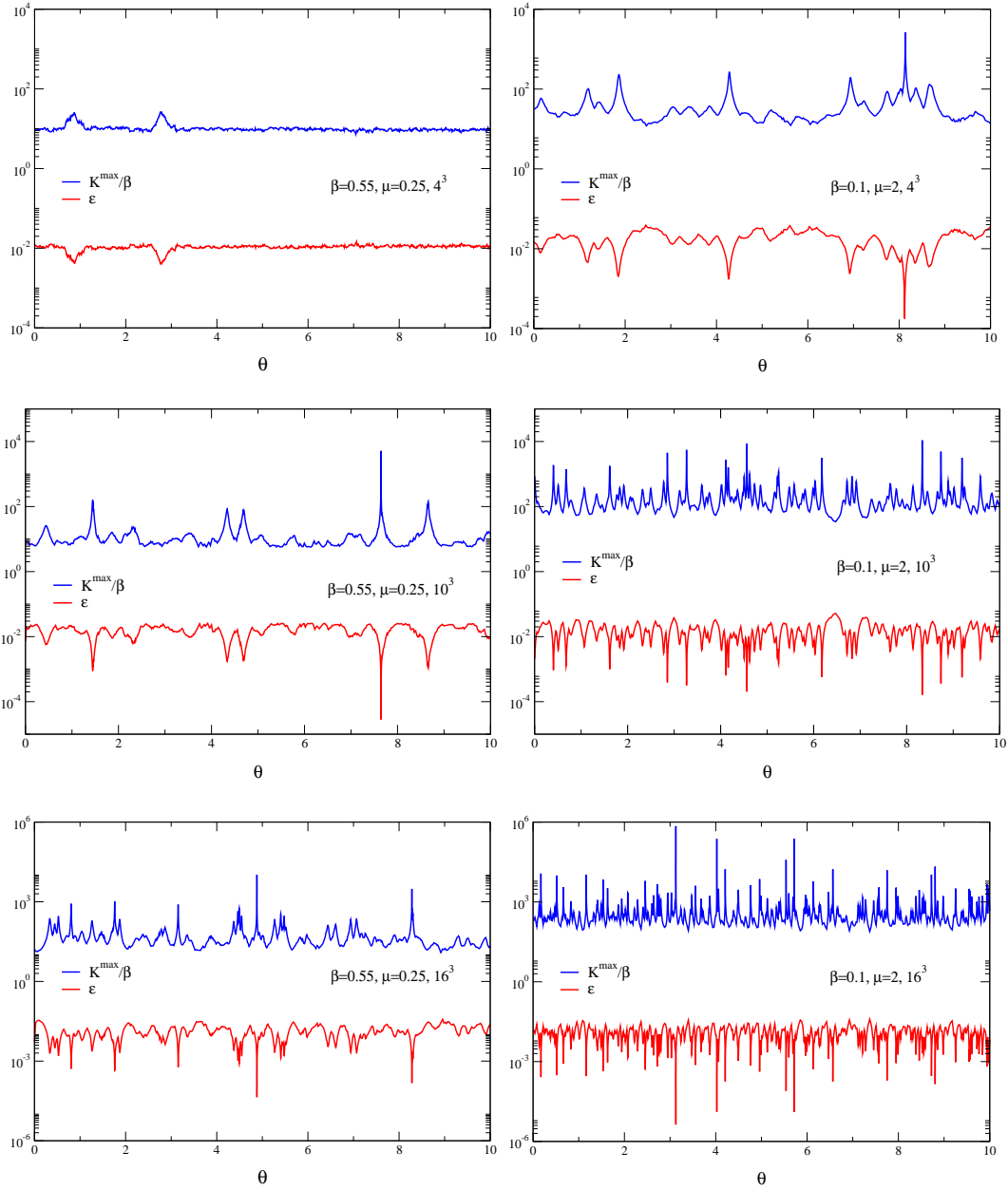

\begin{center}
\epsfig{figure=evol_b0.55_mu0.25_x4.eps,height=5.2cm}
\epsfig{figure=evol_b0.1_mu2_x4.eps,height=5.2cm}
\end{center}
\begin{center}
\epsfig{figure=evol_b0.55_mu0.25_x10.eps,height=5.2cm}
\epsfig{figure=evol_b0.1_mu2_x10.eps,height=5.2cm}
\end{center}
\begin{center}
\epsfig{figure=evol_b0.55_mu0.25_x16.eps,height=5.2cm}
\epsfig{figure=evol_b0.1_mu2_x16.eps,height=5.2cm}
\end{center}
\caption{Example of the Langevin evolution of the maximal drift term 
$K^{\rm max}/\beta$ and the adaptive stepsize $\eps_n$ in the 
three-dimensional XY model with $\beta=0.55$ and $\mu=0.25$ (left) and 
$\beta=0.1$ and $\mu=2$ (right) on lattices of 
size $4^3$ (top), $10^3$ (middle) and  $16^3$ (bottom). The input 
stepsize is $\bar\eps=0.01$. Note the vertical logarithmic scale.}
\label{fig:evol}
\end{figure}

In an attempt to solve these Langevin equations numerically with a fixed 
stepsize, we found that instabilities and runaway trajectories appear so 
frequent, that it is practically impossible to construct a thermalized 
configuration, even when the stepsize is very small, say, $\eps\sim 
10^{-5}$. This becomes worse on larger volumes.\footnote{This is in sharp 
contrast with the relativistic Bose gas in Ref.\ \cite{Aarts:2008wh}, 
where instabilities were not encountered for the parameter values used 
there.} We therefore switch to the adaptive scheme, using the first 
implementation. In Fig.\ \ref{fig:evol} we show examples of the maximal 
drift term $K^{\rm max}$ and the adaptive stepsize $\eps_n$ as a function 
of Langevin time for three different lattice volumes and two values of 
$\beta$ and $\mu$. We observe that the maximal force fluctuates over 
several orders of magnitude during the evolution (note the vertical 
logarithmic scale). Moreover, the frequency and size of the fluctuations 
increase when increasing the lattice volume. This is consistent with the 
picture developed above: on a larger volume there are more opportunities 
to be on a potentially unstable trajectory and subject to large forces. 
The effect also gets worse at larger chemical potential. $K^{\rm max}$ and 
$\eps_n$ are inversely proportional, as expected. We note that although it 
is necessary to use a tiny stepsize occasionally, the algorithm is 
designed such that the evolution will continue with a larger stepsize as 
soon as possible. As a result, the time average of $\eps_n$ is close to 
the input timestep $\bar \eps=0.01$ in all cases.\footnote{In practice, 
one may take $\eps_n\leq \bar\eps$ always, to prevent the appearance of 
large timesteps.}
 We emphasize that after the implementation of this algorithm we 
have not observed any instabilities, for a wide range of parameters 
($0.1<\beta<2$, $0<\mu<6$), lattice sizes (up to $16^3$), and long 
runtimes (we explored millions of timesteps, corresponding to Langevin 
times of several thousand).

 To illustrate the method, we introduce two related models with a real 
action: the XY model with imaginary chemical potential $\mu=i\mu_\rmI$, 
with the action
 \be
 S_{\rm imag} = -\beta\sum_{x,\nu} \cos\left(\phi_{x} - \phi_{x+\hat\nu} +
\mu_\rmI\delta_{\nu,0}\right),
 \ee
 and the phase quenched theory, obtained by taking the absolute value of  
the complex weight, i.e.\ $e^{-S} \to |e^{-S}| \equiv e^{-S_{\rm pq}}$, 
which yields the action
\be
S_{\rm pq} = -\beta\sum_{x,\nu}
 \cos\left(\phi_x-\phi_{x+\hat\nu}\right)
 \cosh\left(\mu\delta_{\nu,0}\right).
\ee
 This is the anisotropic XY model, with direction-dependent 
coupling $\beta_\nu$, where $\beta_0=\beta\cosh\mu$ and $\beta_i=\beta$ 
($i=1,2$). Both models are solved using real Langevin dynamics. The drift 
terms are bounded and there are no instabilities.

\begin{figure}[t]
\begin{center}
\epsfig{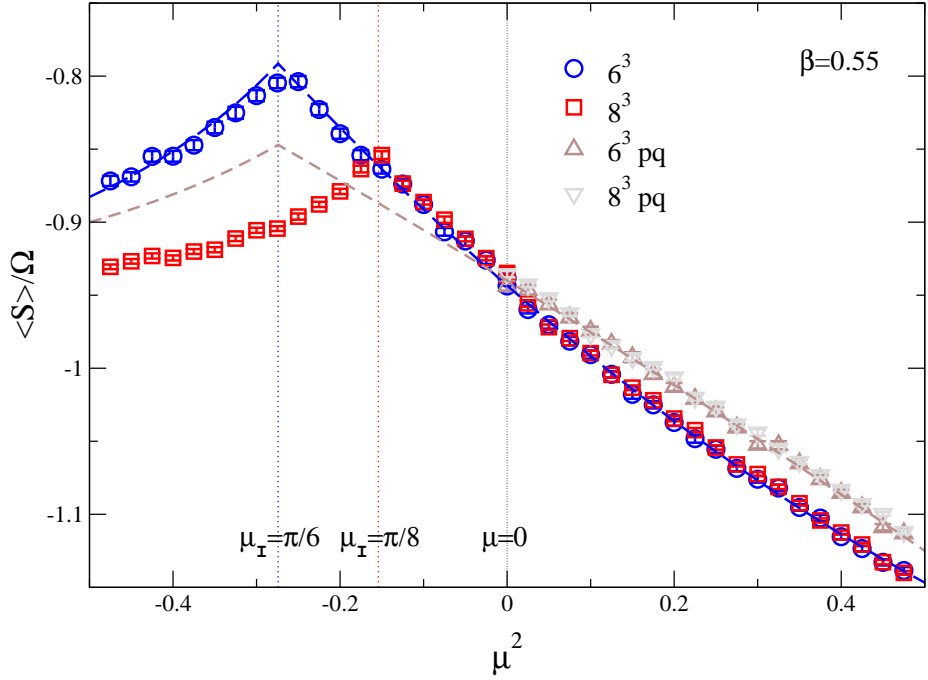}
\end{center}
  \caption{Action density $\bra S\ket/\Omega$ in the three-dimensional XY 
model as a function of $\mu^2$ at $\beta=0.55$ on lattices of size $6^3$ 
and $8^3$, for the full theory (circle, square, $\mu^2>0$), at imaginary 
$\mu$ (circle, square, $\mu^2<0$), and phase quenched (triangles, 
$\mu^2>0$). The vertical lines at $\mu_\rmI=\pi/N_\tau$ indicate the 
Roberge-Weiss lines at imaginary $\mu$. The dashed lines are the 
second-order fits (\ref{eq:fit1}, \ref{eq:fit2}), incorporating the RW 
reflection symmetry.
 }
 \label{fig:action}
\end{figure}

In Fig.\ \ref{fig:action} we show the expectation value of the action 
density $\bra S\ket/\Omega$ in the high-$\beta$ phase, at $\beta=0.55$, 
for small values of $\mu^2$.\footnote{Recall that at zero chemical 
potential, the three-dimensional XY model has a continuous phase 
transition at $\beta_c(\mu=0) \approx 0.4542$ (see e.g.\ Ref.\ 
\cite{Campostrini:2000iw}), separating the symmetric phase at small 
$\beta$ from the symmetry broken phase at large $\beta$.} The lattice 
sizes are $6^3$ and $8^3$, showing that finite size effects are under 
control. The result at imaginary $\mu$ appears at $\mu^2<0$, while the 
full and phase quenched results are plotted at $\mu^2>0$. At $\mu=0$ all 
results agree (within the statistical error). The full and phase quenched 
results differ, as can be expected from e.g.\ a Taylor expansion of the 
observable for small $\mu$.  The results for imaginary and real $\mu$ 
appear continuous around $\mu^2=0$, which is expected from the analyticity 
of the partition function in $\mu^2$ on a finite lattice. The lines 
indicate second-order fits to the data on the $6^3$ lattice, with the 
results
 \bea
\label{eq:fit1}
\bra S\ket/\Omega =&&\hm -0.9433(7) - 0.502(4) \mu^2 + 
0.19(1) \mu^4,
\\
\label{eq:fit2}
\bra S\ket_{\rm pq}/\Omega =&&\hm -0.940(2)\,\,\, - 0.35(2) \mu^2 \,\,\,
- 0.04(3) \mu^4.
 \eea
 In the first case, the data at real and imaginary $\mu$ are combined in 
the fit.


 For imaginary $\mu$ we observe a cusp at $\mu_\rmI=\pi/N_\tau$. This is 
similar to the Roberge-Weiss transition in QCD \cite{Roberge:1986mm,de 
Forcrand:2002ci} and is due to the periodicity $\mu_\rmI \to\mu_\rmI+ 
2\pi/N_\tau$.\footnote{One way to see this is by using a field 
redefinition, $\phi_x = \phi_x' +\mu_\rmI \tau$, which moves the $\mu$ 
dependence to the boundary condition $\phi_{N_\tau,\vecx}' = 
\phi_{0,\vecx}'-\mu_\rmI N_\tau$, similar as in fermionic models. The 
center symmetry is of course trivial in this model.} 
Note that the dashed lines in Fig.\ \ref{fig:action} reflect this 
symmetry. 
It would therefore be 
interesting to determine the phase structure of the XY model at imaginary 
chemical potential and finite $N_\tau$.
 In the three-dimensional thermodynamic limit ($N_\tau$ is taken to 
infinity as well) and vanishing chemical potential, the magnetized and 
symmetric phase are separated at the critical coupling 
$\beta_c(\mu=0)$. Consequently it is intriguing to analyse 
the interplay between the putative Roberge-Weiss transition and the 
standard magnetization transition in this limit, in particular since it 
would result in a breakdown of analyticity of $\beta_c(\mu^2)$ around 
$\mu^2=0$ and make $\beta_c(\mu=0)$ a multicritical point.

\section{Heavy dense limit of QCD}
\label{sec:hdm}
\setcounter{equation}{0}

To show the generality of the adaptive stepsize method we now apply it to 
the heavy dense limit of QCD in four dimensions. Here we shall present 
results obtained with the second algorithm. The stochastic quantization 
for this theory was studied  
in Ref.\ \cite{Aarts:2008rr}. Here we briefly 
repeat the essential equations; we refer to Ref.\ \cite{Aarts:2008rr} for 
further details.

 The gluonic part of the action is the standard Wilson SU(3) action,
 \be
S_B[U] = -\beta \sum_x
\mathop{\sum_{\mu,\nu}}_{\mu<\nu}
\left(\frac{1}{6}\left[ \Tr
U_{x,\mu\nu}+ \Tr U_{x,\mu\nu}^{-1}\right]-1\right),
\ee
where $U_{x,\mu\nu}$ are the plaquettes and $\beta = 6/g^2$.
 The fermion determinant (starting from Wilson fermions) is approximated 
as
\be
\det M \approx \prod_{\xv}
\det\left( 1 + h e^{\mu/T} \cP_{\xv} \right)^2
\det\left( 1 + h e^{-\mu/T} \cP_{\xv}^{-1} \right)^2,
\label{eq:det}
\ee
where $h=(2\kappa)^{N_\tau}$. Here $\kappa$ is the Wilson hopping 
parameter 
and $N_\tau=1/T$ the number of sites in the temporal direction. The 
lattice spacing $a\equiv 1$.  The determinant refers to colour space 
only. The (conjugate) Polyakov loops are
\be
 \cP_\xv = \prod_{\tau=0}^{N_\tau-1} U_{(\tau,\xv),4},
\;\;\;\;\;\;\;\;\;\;\;\;\;\;\;\;
 \cP_\xv^{-1} = \prod_{\tau=N_\tau-1}^{0} U_{(\tau,\xv),4}^{-1}.
\ee
 A formal derivation of Eq.\ (\ref{eq:det}) follows by considering the 
heavy ($\kappa\to 0$) and dense ($\mu\to\infty$) limit, keeping the 
product $\kappa e^\mu$ fixed (see Ref.\ \cite{DePietri:2007ak} and 
references therein). The anti-quark contribution is kept to preserve the 
symmetry under complex conjugation.

The Langevin process is
\be
 \label{eq:flang}
 U_{x,\mu}'  = R_{x,\mu}\, U_{x,\mu},
\;\;\;\;\;\;\;\;\;\;\;\;
 R_{x,\mu} = \exp\left[i\lambda_a\left(\eps K_{x\mu a} +\sqrt{\eps}
\eta_{x\mu a} \right)\right],
\ee
where $\lambda_a$ are the Gell-Mann matrices and the noise satisfies
$\bra \eta_{x\mu a}\ket = 0$, 
$\bra \eta_{x\mu a}\eta_{y\nu b} \ket = 2\delta_{\mu\nu}\delta_{ab}\delta_{xy}$.
 The drift term $K_{x\mu a} = -D_{x\mu a}\left( S_B +S_F \right)$, where 
$S_F = -\ln\det M$, is complex due to the fermion contribution. For 
explicit expressions, see Ref.\ \cite{Aarts:2008rr}.

\begin{figure}[t]
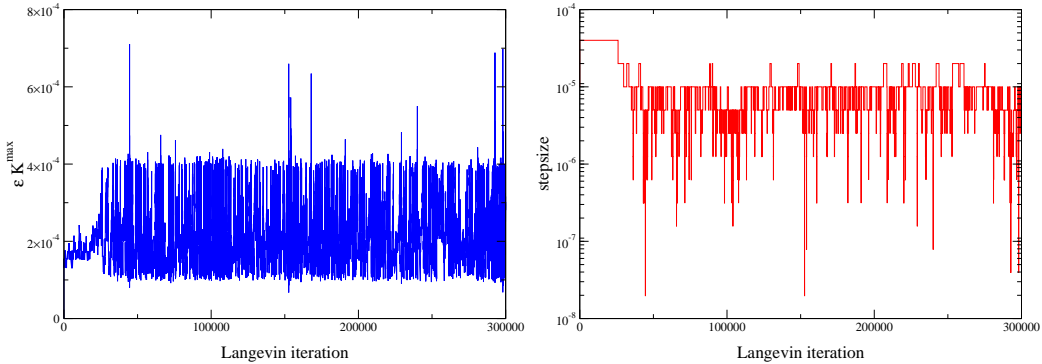

\begin{center}
\epsfig{figure=fig_kmax_HDM.eps,height=4.8cm}
\epsfig{figure=fig_stepsize_HDM.eps,height=4.8cm}
\end{center}
\caption{Example of the Langevin evolution of the maximal drift term 
$\eps K^{\rm max}$ (left) and the adaptive stepsize $\eps$ (right)
in the heavy dense limit of QCD with $\beta=5$, $\kappa=0.12$ and 
$\mu=0.7$ on a  lattice of size $2^4$, using ${\cal K}=2\times 10^{-4}$.
}
\label{fig:hdm1}
\end{figure}

\begin{figure}[t]
\begin{center}
\epsfig{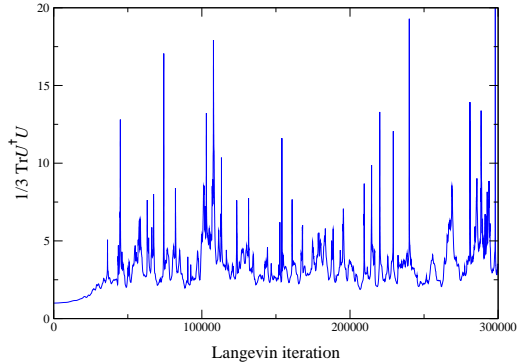}
\end{center}
\caption{As above for $\frac{1}{3}\Tr U_4U_4^{\dag}$, indicating the 
deviation from unitarity during the evolution.
}
\label{fig:hdm2}
\end{figure}

This Langevin process suffers from instabilities, which can be partially 
controlled using small enough stepsize. Applying the second formulation of 
the adaptive stepsize algorithm completely eliminates runaways. In this 
setup measurements are performed after a Langevin time interval of length 
$\Delta \vartheta$ (this defines one ``iteration").  The number of sweeps 
$N_{\rm sweep}$ in an iteration depends on the Langevin timestep 
$\epsilon$: if the latter is decreased by a factor $p$, $N_{\rm sweep}$ is 
increased by the same factor, and conversely. This makes the statistical 
analysis straightforward, since each iteration has the same weight
$\Delta \vartheta = \eps\, N_{\rm sweep}$. Note 
that $N_{\rm sweep}$ will no longer be decreased and correspondingly 
$\epsilon$ will no longer be increased if the former reaches~1. 

To demonstrate this approach, we show in Fig.\ \ref{fig:hdm1} a 
characteristic evolution of $\eps K^{\rm max}$ (left) and the stepsize 
$\epsilon$ (right), using ${\cal K}=2\times 10^{-4}$ and $p=2$, with 
initial $\epsilon=1\times 10^{-5}$, $N_{\rm sweep}=16$. The total number 
of iterations is $N_{\rm iter}=3\times 10^5$ leading to a total Langevin 
time $\vartheta_{\rm tot} = N_{\rm iter}\times \Delta \vartheta = 48$. 
Varying 
these input parameters by factors of two or more does not change the 
results but may only affect the statistics. $K^{\rm max}$ is the maximum 
value of $|K_{x\mu a}|$ over $x$, $\mu$ and $a$ in the last sweep of an 
iteration. We observe that the product $\epsilon K^{\rm max}$ remains 
bounded, as required, while the stepsize (and hence $K^{\rm max}$) 
fluctuate substantially. In Fig.\ \ref{fig:hdm2} we show the evolution of 
$\frac{1}{3}\Tr U_4 U_4^{\dag}$, which measures the deviation from 
unitarity \cite{Aarts:2008rr}, acknowledging the typical fluctuations in 
stationary regime.

 As mentioned above, after the implementation of this algorithm, we have 
not encountered any instability.

\section{Conclusion}
\label{sec:conclusion}
\setcounter{equation}{0}

Instabilities can make complex Langevin simulations extremely 
problematic. We have shown that they result from large fluctuations in 
the magnitude of the drift term and the presence of unstable directions 
in complexified field space. This can be cured by using an adaptive 
time-local stepsize. The scheme is generic and could be applied to 
other theories such as QCD.
 While we have no proof that stable evolution is guaranteed, we have not 
encountered any instability using this method in the XY model at finite 
chemical potential and QCD in the heavy dense limit, for a wide selection 
of parameter values and system sizes. With a fixed stepsize on the other 
hand, instabilities can be so frequent that it is virtually impossible to 
generate a thermalized ensemble.

Runaways are due to specific instabilities of the complexified Langevin equations 
caused by the strong increase of the drift in the non-compact imaginary directions. 
We have shown that these runaways can be eliminated by using a dynamical step size, 
which indicates that they are not a question of principle for complex Langevin 
dynamics but one of numerical accuracy in following the trajectories. We consider 
this to be an important result of our analysis. Of course, for practical calculations 
one should consider further optimization of the algorithms by applying methods 
developed for general real Langevin processes (see, e.g., Ref.\ 
\cite{Petersen:1996by}) after analysing their adequacy to the complex Langevin 
problems of interest.

We emphasize that the adaptive stepsize permits a fine tracing of the 
drift trajectories. If the process picks up a diverging trajectory the 
noise term will typically kick the process off it. The present results 
suggest therefore that runaways are not due to following diverging 
trajectories \cite{Aarts:2009uq} but rather due to following ``wrong" 
trajectories, i.e.\ trajectories which, because of accumulating errors in 
the evaluation of the drift, do not belong to the dynamics of the problem. 
This both stresses the necessity of ensuring precision in the calculation 
and helps in disentangling various sources of deficiency in the search for 
a reliable method.


 \vspace*{0.5cm}
 \noindent
 {\bf Acknowledgments.}

We thank Simon Hands, Owe Philipsen and Denes Sexty for discussions. 
I.-O.S. thanks the MPI for Physics M\"unchen and Swansea 
University for hospitality. G.A.\ and F.J.\ thank the Blue C Facility at 
Swansea University for computational resources. G.A.\ and F.J.\ are 
supported by STFC.


\begin{thebibliography}{10}

\bibitem{Parisi:1980ys}
  G.~Parisi and Y.~s.~Wu,
  Sci.\ Sin.\  {\bf 24} (1981) 483.

\bibitem{Parisi:1984cs}
  G.~Parisi,
  Phys.\ Lett.\  B {\bf 131} (1983) 393.

\bibitem{Klauder:1983}
 J.~R.~Klauder,
 Stochastic quantization, 
 in: H.~Mitter, C.B.~Lang (Eds.), Recent Developments in High-Energy 
Physics, Springer-Verlag, Wien, 1983, p.\ 351.

\bibitem{Damgaard:1987rr}
  P.~H.~Damgaard and H.~H\"uffel,
  Phys.\ Rept.\  {\bf 152} (1987) 227.



\bibitem{Karsch:1985cb}
  F.~Karsch and H.~W.~Wyld,
  Phys.\ Rev.\ Lett.\  {\bf 55} (1985) 2242.

\bibitem{Aarts:2008rr}
  G.~Aarts and I.-O.~Stamatescu,
  JHEP {\bf 0809} (2008) 018
  [0807.1597 [hep-lat]].

\bibitem{Aarts:2008wh}
  G.~Aarts,
  Phys.\ Rev.\ Lett.\  {\bf 102} (2009) 131601
  [0810.2089 [hep-lat]].

\bibitem{Aarts:2009hn}
  G.~Aarts,
  JHEP {\bf 0905} (2009) 052
  [0902.4686 [hep-lat]].


                                                                                
\bibitem{Berges:2005yt}
  J.~Berges and I.-O.~Stamatescu,
  Phys.\ Rev.\ Lett.\  {\bf 95} (2005) 202003
  [hep-lat/0508030].
                                                                                
\bibitem{Berges:2006xc}
  J.~Berges, S.~Borsanyi, D.~Sexty and I.~O.~Stamatescu,
  Phys.\ Rev.\  D {\bf 75} (2007) 045007
  [hep-lat/0609058].
                                                                                
\bibitem{Berges:2007nr}
  J.~Berges and D.~Sexty,
  Nucl.\ Phys.\  B {\bf 799} (2008) 306
  [0708.0779 [hep-lat]].



\bibitem{Bernard:2001wh}
  C.~W.~Bernard and V.~M.~Savage,
  Phys.\ Rev.\  D {\bf 64} (2001) 085010
  [hep-lat/0106009].

                                                                                
\bibitem{Pehlevan:2007eq}
  C.~Pehlevan and G.~Guralnik,
  Nucl.\ Phys.\  B {\bf 811}, 519 (2009)
  [0710.3756 [hep-th]],
  Nucl.\ Phys.\  B {\bf 822} (2009) 349
  [0902.1503 [hep-lat]].



\bibitem{Klauder:1985b}
   J.~R.~Klauder and W.~P.~Petersen,
   J.\ Stat.\ Phys.\ 39 (1985) 53.


\bibitem{Ambjorn:1985iw}
  J.~Ambjorn and S.~K.~Yang,
  Phys.\ Lett.\  B {\bf 165} (1985) 140.


\bibitem{Ambjorn:1986fz}
  J.~Ambjorn, M.~Flensburg and C.~Peterson,
  Nucl.\ Phys.\  B {\bf 275} (1986) 375.


\bibitem{Chandrasekharan:2008gp}
  S.~Chandrasekharan,
  PoS {\bf LATTICE2008} (2008) 003
  [0810.2419 [hep-lat]].

\bibitem{Hasenfratz:1983ba}
  P.~Hasenfratz and F.~Karsch,
  Phys.\ Lett.\  B {\bf 125} (1983) 308.

\bibitem{Endres:2006xu}
  M.~G.~Endres,
  Phys.\ Rev.\  D {\bf 75} (2007) 065012
  [hep-lat/0610029].


\bibitem{Campostrini:2000iw}
  M.~Campostrini, M.~Hasenbusch, A.~Pelissetto, P.~Rossi and E.~Vicari,
  Phys.\ Rev.\  B {\bf 63} (2001) 214503
  [cond-mat/0010360].


\bibitem{Roberge:1986mm}
  A.~Roberge and N.~Weiss,
  Nucl.\ Phys.\  B {\bf 275}, 734 (1986).

\bibitem{de Forcrand:2002ci}
  P.~de Forcrand and O.~Philipsen,
  Nucl.\ Phys.\  B {\bf 642} (2002) 290
  [hep-lat/0205016].


\bibitem{DePietri:2007ak}
  R.~De Pietri, A.~Feo, E.~Seiler and I.~O.~Stamatescu,
  Phys.\ Rev.\  D {\bf 76} (2007) 114501
  [0705.3420 [hep-lat]].

\bibitem{Petersen:1996by}
  W.~P.~Petersen,
  hep-lat/9602008.

\bibitem{Aarts:2009uq}
  G.~Aarts, E.~Seiler and I.~O.~Stamatescu,
  Phys.\ Rev.\ D (to appear) [0912.3360 [hep-lat]].



\end{thebibliography}
\end{document}